\documentclass[aps,prx,reprint,preprintnumbers,superscriptaddress,nofootinbib,longbibliography,floatfix]{revtex4-1}

\usepackage[T1]{fontenc}
\usepackage{booktabs}
\usepackage{graphicx}
\usepackage{xspace}
\usepackage{multirow}
\usepackage{amssymb,amsmath}
\usepackage{mathtools}
\usepackage{xstring}
\usepackage[dvipsnames]{xcolor}
\usepackage{afterpage}
\usepackage{hyperref}
\hypersetup{
  colorlinks=true,
  citecolor=blue,
  linkcolor=blue,
  urlcolor=blue
}
\usepackage{cleveref}

\DeclareRobustCommand{\ccite}[1]{\IfSubStr{#1}{,}{refs.~}{ref.~}\cite{#1}}
\DeclareRobustCommand{\Ccite}[1]{\IfSubStr{#1}{,}{Refs.~}{ref.~}\cite{#1}}

\newcommand{\eg}{\textit{e.g.}\xspace}

\newcommand{\GeV}{\ensuremath{\text{\,Ge\kern -0.1em V}}\xspace}
\newcommand{\TeV}{\ensuremath{\text{\,Te\kern -0.1em V}}\xspace}

\newcommand{\llh}[1][\text{NF}]{{\ensuremath{\mathcal{L}_{#1}}}\xspace}

\newcommand{\by}{\ensuremath{\boldsymbol{y}}\xspace}
\newcommand{\bysim}[1][]{\ensuremath{\by_{\text{sim}#1}}\xspace}
\newcommand{\byexp}[1][]{\ensuremath{\by_{\text{exp}#1}}\xspace}

\newcommand{\mmp}[1]{\,#1}

\newcommand{\emd}{\ensuremath{\text{EMD}}\xspace}

\definecolor{Red}{rgb}{1.,0.,0.}
\definecolor{Grn}{rgb}{0.,0.75,0.}
\definecolor{Blu}{rgb}{0.,0.,1.}
\definecolor{Purp}{rgb}{0.3,0.1,0.6}
\definecolor{Mag}{rgb}{1.,0.,1.}
\definecolor{Cor}{rgb}{1.,0.5,0.5}
\definecolor{BN}{rgb}{0.1,0.5,0.2}
\definecolor{AS}{rgb}{1.0, 0.6, 0.4}

\newcommand{\reportnumber}{FERMILAB-PUB-24-0784-CSAID, MCNET-24-18}
\usepackage[angle=0,scale=1,color=black,firstpage=true,opacity=1]{background}
\SetBgContents{\footnotesize\sf{\reportnumber}}
\SetBgPosition{current page.north east}
\SetBgHshift{-2in}
\SetBgVshift{-0.5in}

\begin{document}
\title{Rejection Sampling with Autodifferentiation\\
\small Case study: Fitting a Hadronization Model}
\author{Nick Heller}
\email{nheller@berkeley.edu}
\affiliation{Lawrence Berkeley National Laboratory, Berkeley, CA 94720, USA}

\author{Phil Ilten}
\email{philten@cern.ch}
\affiliation{Department of Physics, University of Cincinnati, Cincinnati, Ohio 45221, USA}

\author{Tony Menzo}
\email{menzoad@mail.uc.edu}
\affiliation{Department of Physics, University of Cincinnati, Cincinnati, Ohio 45221, USA}

\author{Stephen Mrenna}
\email{mrenna@fnal.gov}
\affiliation{Scientific Computing Division, Fermilab, Batavia, Illinois 60510, USA}
\affiliation{Department of Physics, University of Cincinnati, Cincinnati, Ohio 45221, USA}

\author{Benjamin Nachman}
\email{bpnachman@lbl.gov}
\affiliation{Lawrence Berkeley National Laboratory, Berkeley, CA 94720, USA}

\author{Andrzej Siodmok}
\email{andrzej.siodmok@uj.edu.pl}
\affiliation{Jagiellonian University, Łojasiewicza 11, 30-348 Kraków, Poland}

\author{Manuel Szewc}
\email{szewcml@ucmail.uc.edu}
\affiliation{Department of Physics, University of Cincinnati, Cincinnati, Ohio 45221, USA}
\affiliation{International Center for Advanced Studies (ICAS), ICIFI and ECyT-UNSAM, 25 de Mayo y Francia, (1650) San Mart\'{i}n, Buenos Aires, Argentina}

\author{Ahmed Youssef}
\email{youssead@ucmail.uc.edu}
\affiliation{Department of Physics, University of Cincinnati, Cincinnati, Ohio 45221, USA}

\begin{abstract}
We present an autodifferentiable rejection sampling algorithm termed \textit{Rejection Sampling with Autodifferentiation} (RSA). In conjunction with reweighting, we show that RSA can be used for efficient parameter estimation and model exploration. Additionally, this approach facilitates the use of unbinned machine-learning-based observables, allowing for more precise, data-driven fits. To showcase these capabilities, we apply an RSA-based parameter fit to a simplified hadronization model.
\end{abstract}

\maketitle

\section{Introduction}\label{sec:introduction}

Simulations are essential tools for connecting fundamental theories with observations in many physical sciences.  These simulations are built from physical and phenomenological models containing many parameters.  In nearly every analysis of experimental data in particle and nuclear physics, there is a need to produce simulated datasets with one or more of the simulation parameters varied.  Traditionally, varying the parameters requires generating new, statistically independent synthetic datasets.  Fitting the parameters further requires interpolating between such datasets and this has only been possible by first summarizing the data into a small number of observables.  Therefore, the current approaches are unable to use all of the available information and are computationally expensive.

One solution to these challenges is to build sophisticated surrogate models.  Modern machine learning (ML), and generative modeling in particular, have introduced a number of tools that can accurately and rapidly emulate physics-based simulations.  A key benefit of neural network-based surrogate models is that they are inherently \textit{autodifferentiable} (AD).  In the context of parameter fitting, derivatives are computed with respect to statistics comparing data and simulation (e.g. the $\chi^2$ difference of two histograms).  Gradient descent is an efficient approach to picking the model parameters that best describe the data.  The \textit{auto} in autodifferentiable refers to the ability to compute gradients accurately (machine-precision limited) and efficiently. A number of recent studies have shown how to fit a ML-based surrogate model to data using such a framework~\cite{Ilten:2022jfm,Ghosh:2022zdz,Chan:2023ume,Bierlich:2023zzd,Chan:2023icm,Bierlich:2024xzg}.

While highly flexible, ML surrogate models can also be unwieldy due to their large number of parameters.  Alternatively, \textit{differentiable simulations} benefit from both the physical structure of classical simulations and the AD of ML surrogate models.   If the simulation itself is written in an AD-compatible programming language, then it could be possible to compute derivatives of statistics through the simulation itself.  With far fewer parameters than neural networks, such an approach may be more effective than using a generic deep generative model.  There have been a number of proposals for differentiable physics simulators in particle and nuclear physics, including for matrix element calculations~\cite{Carrazza:2021gpx,Heinrich:2022xfa,Heimel:2024wph}, parton showers~\cite{Nachman:2022jbj}, and detector simulations~\cite{MODE:2022znx,aehle2024optimizationusingpathwisealgorithmic,Cheong:2022zov,Strong:2023oew,Gasiorowski:2023tqf}.  

An essential incompatibility between AD and Monte Carlo (MC) simulations is that the latter contains many operations that are not differentiable\footnote{See ref.~\cite{Kagan:2023gxz} for some cases where approximations are possible.}.  One widely-used technique in MC simulations is rejection sampling, which is a general algorithm for sampling from a probability density whose normalization is not known.  By construction, a sample is accepted or rejected based on a (pseudo-) random number.  Changes to the model parameters change the accept-reject threshold value.  The derivative of the samples with respect to the underlying parameter is almost everywhere zero except at one value, where it is infinity.  

We propose a new technique that allows for the autodifferentiation of rejection sampling without making any approximations.   The key idea is to use event weights that are differentiable.  Reweighting MC event samples is a powerful technology for the efficient exploration of model parameter-space \cite{Gainer:2014bta}. An efficient algorithm, based on standard accept-reject sampling, for computing event weights in the absence of an explicit normalized probability density function was introduced in refs.~\cite{Mrenna:2016sih,Bellm:2016voq} and used to provide uncertainty estimation associated with different parton shower parameterizations. A similar procedure was recently developed in ref.~\cite{Bierlich:2023fmh} to provide hadronization uncertainty estimation. In this work, we employ this idea to build an automated black-box parameter tuning referred to from here on as \textit{Rejection Sampling with Autodifferentiation} (RSA).  To demonstrate this method, we show how to fit the parameters of a hadronization model, which has gained recent attention in the context of ML surrogate models~\cite{Ilten:2022jfm,Ghosh:2022zdz,Chan:2023ume,Bierlich:2023zzd,Chan:2023icm,Bierlich:2024xzg}.  An AD hadronization model is an alternative approach to generic neural networks that builds in physical priors in order to significantly reduce the number of parameters needed to describe the data.  Along the way, we explore new ways of building test statistics for the parameter fitting using tools from optimal transport.

The paper is structured as follows: In Sec.~\ref{sec:rejection_sampling} we introduce rejection sampling and the RSA methodology, in Sec.~\ref{sec:tuning_w_RSA} we outline the RSA-based algorithm for parameter estimation and introduce the problem in the context of hadronization models, in Sec.~\ref{sec:data} we describe the data generation pipeline, in Sec.~\ref{sec:results} we summarize our results and finally in Sec.~\ref{sec:conclusion} we conclude with final remarks and highlight future work.

\section{Rejection sampling with autodifferentiation}\label{sec:rejection_sampling}
\subsection{Accept-reject sampling}
Consider a random variable $x$ over the domain $\mathcal{X}$ with a probability density $P(x;\boldsymbol{\theta})$ known up to a normalizing constant with $\mathcal{N}$ tuneable parameters $\boldsymbol{\theta} \equiv \{\theta_1, \cdots, \theta_\mathcal{N}\}$.  Assume the maximum $\hat{P}$ is known (or can be estimated numerically) such that $P(x) \leq \hat{P}$ for all $x \in \mathcal{X}$.
Samples of $x$ can be obtained by:
\begin{enumerate}
  \item Uniformly sampling a trial value $x_1 \in \mathcal{X}$.
  \item Computing the acceptance probability 
        \begin{equation}
          P^1_{\text{accept}}(x_1;\boldsymbol{\theta}) = \frac{P(x_1; \boldsymbol{\theta})}{\hat{P}}.
        \end{equation}
  \item Sampling an additional uniformly distributed value $s_1 \in [0, \hat{P}]$.
  \item If $P^1_{\text{accept}} (x_1) \leq s_1$ accept the trial $x_1$, otherwise, reject and return to step 1. Repeat $n$-trials until $P^n(x_n) \leq s_n$ i.e.~until a trial is accepted.
\end{enumerate}
Note that $\hat{P}$ is unbounded from above, allowing the efficiency of the accept-reject algorithm to be tuned, with values larger than the minimum $\hat{P}_{\text{min}} \equiv \text{max}[P(x)]$ decreasing the efficiency.
A single element of an accept-reject sampling will have one accepted value $x_{\text{accept}}$ and $n-1$ rejected trials $\boldsymbol{x}_{\text{reject}} \equiv \{x^1_{\text{reject}}, \cdots, x^{{n-1}}_{\text{reject}}\}$.

\subsection{Alternative accept-reject sampling}
Given a sample $\boldsymbol{x}$, a different sample using the same probability model but with different parameters $\boldsymbol{\theta}' \neq \boldsymbol{\theta}$ can be obtained by assigning each trial $x_i \in \boldsymbol{x}$ a statistical weight \cite{Mrenna:2016sih}.
 The total weight of each element is obtained by assigning the accepted value the relative acceptance probability
\begin{equation}
  w_{\text{accept}} \equiv \frac{P_{\text{accept}}(x_{\text{accept}};\boldsymbol{\theta}')}{P_{\text{accept}}(x_{\text{accept}};\boldsymbol{\theta})} = \frac{P(x_\text{accept}; \boldsymbol{\theta}')}{P(x_\text{accept}; \boldsymbol{\theta})}\,,
\end{equation}
and each rejected trial value the relative rejection probability 
\begin{equation}
  w^{j}_{\text{reject}} \equiv \frac{1-P_{\text{accept}}(x^j_{\text{reject}}, \boldsymbol{\theta}')}{1 - P_{\text{accept}}(x^j_{\text{reject}}, \boldsymbol{\theta})} = \frac{\hat{P} - P(x^j_{\text{reject}}, \boldsymbol{\theta}')}{\hat{P} - P(x^j_\text{reject}, \boldsymbol{\theta})}.
\end{equation}
The full weight, $w$, of an element in the sample is given by
\begin{equation}
  w = w_{\text{accept}} \prod_{j = 1}^{n-1} w^{j}_{\text{reject}}.
\end{equation}
These weights encode the likelihood ratio between two alternative parameterizations of the same distribution with $w < 1$ and $w > 1$ implying the sample is less or more probable in the new parameterization $\boldsymbol{\theta}'$. 

Assuming that $P(x)$ is smooth, the above prescription is inherently differentiable and facilitates the computation of \textit{weight gradients} with respect to $\boldsymbol{\theta}'$. If $P(x)$ is sampled repeatedly within a larger simulation pipeline, we may also compute gradients of simulation outputs (which may be arbitrarily complicated functions of the sampled values of $x$) via weighted binned or \textit{unbinned} distributions. In practice, this requires the ability to efficiently compute and store numerical gradients. Modern machine learning requirements meet this demand with well-established, GPU-accelerated, differential programming libraries, such as PyTorch \cite{PyTorch} (used in this paper), providing powerful numerical autodifferentiation engines that allow for efficient, out-of-place gradient computations. The parameters of interest can be implemented as differentiable (learnable) objects within custom classes (such as a PyTorch \texttt{Module}) whose forward function facilitates the extraction of weights for a given batch of training data. We refer to this melding as \textit{Rejection Sampling with Autodifferentiation} (RSA).

\section{Parameter estimation with RSA}\label{sec:tuning_w_RSA}
A practical use-case for our studies is to consider the fitting or tuning of parameters inside of Monte Carlo event generators~\cite{Bierlich:2022pfr,Bellm:2015jjp,Sherpa:2019gpd}.
Tuning consists of simulating a large number of events with a `base' parameterization $\boldsymbol{\theta}$, comparing the output distributions to experimental data, and updating the base parameterization to a new parameterization $\boldsymbol{\theta}'$ following a prescribed procedure \cite{Skands:2009zm,Buckley:2009bj,Skands:2014pea,Ilten:2016csi,Krishnamoorthy:2021nwv}. The first and third steps present as the primary bottlenecks encountered when performing full event-generator tunes. The generation of simulated events at each parameter point can take $\mathcal{O}(\text{hours/million events})$ of CPU time. This makes even a modest tuning exercise (a few parameters scanned over a hypercube consisting of 5--10 parameter values per dimension) computationally prohibitive. The computational demand can be further compounded by the non-trivial task of choosing an updated parameterization, with inefficient choices significantly increasing computation time. Below, we describe a novel tuning paradigm that alleviates this computational burden through statistical reweighting\footnote{For a detailed discussion and explicit timing comparisons between reweighted and re-simulated events we refer the reader to ref.~\cite{Bierlich:2023fmh}.} and utilizes modern autodifferention and optimization paradigms to choose updated parameterizations. This approach is similar but distinct to that of DCTR \cite{Andreassen:2019nnm} which also performs event reweighting and parameter tuning, but utilizes a deep neural network classifier (instead of the analytic unnormalized density function) to extract event weights and perform parameter tuning. Other machine-learning-inspired tuning paradigms, such as those described in \cite{Lazzarin:2020uvv}, do not utilize reweighting. Below we will illustrate the efficacy of parameter estimation using RSA in the context of fitting a hadronization model.

\subsection{Hadronization Models}\label{subsec:hadronization}

\begin{figure*}[t!]
  \centering
  \includegraphics[width=0.85\linewidth]{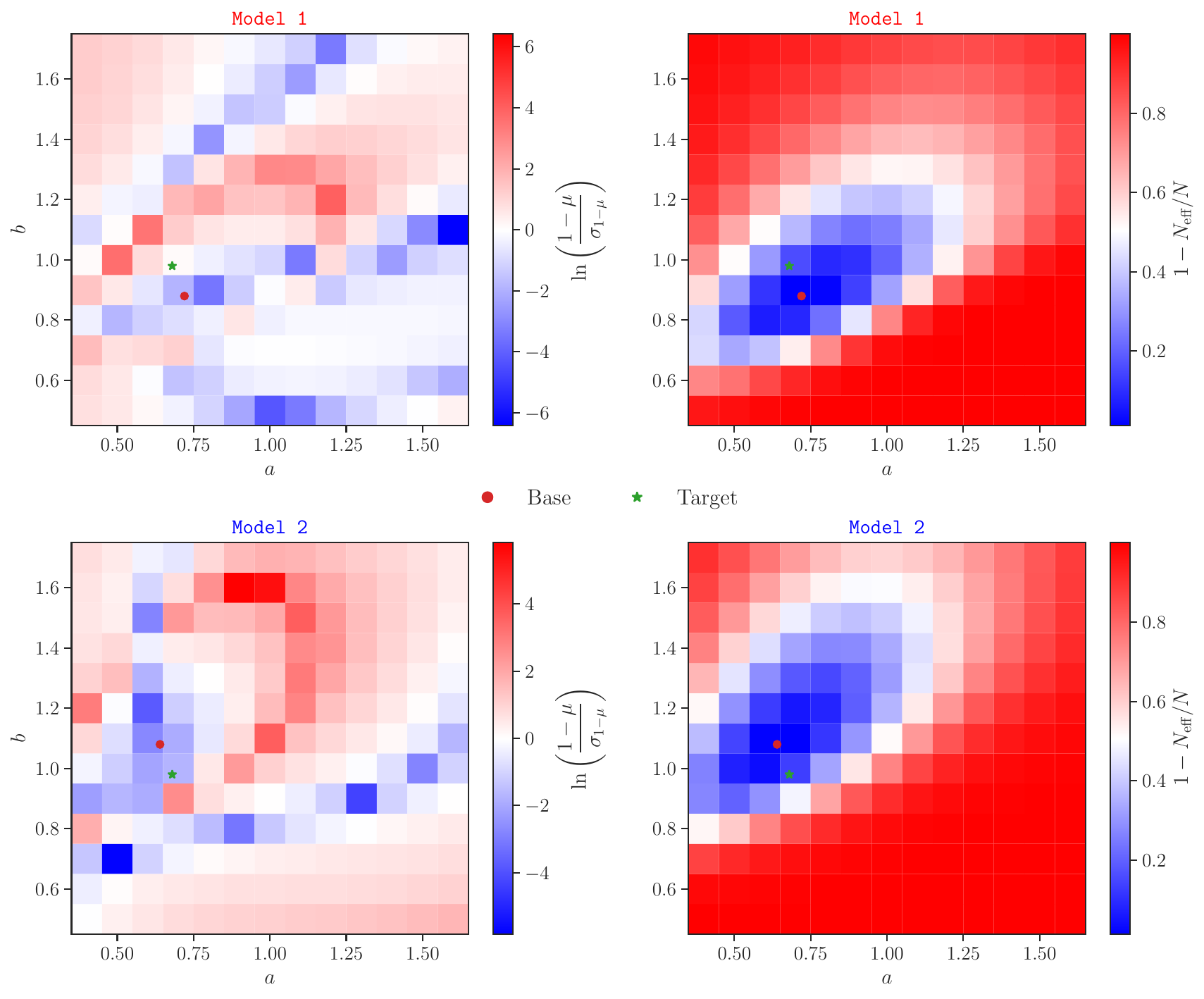}
  \caption{The reweighting metrics defined in eq.~\ref{eq:metrics} over the $(a,b)$ parameter plane for the \textcolor{red}{\texttt{Model 1}} and \textcolor{blue}{\texttt{Model 2}} base parameterizations denoted by the red dots. The green star denotes the `experimental' or target parameterization. For both metrics, values closer to zero imply better coverage and effective statistics. 
  }
  \label{fig:a_b_reweighting_metrics}
\end{figure*}

Hadronization models are used to connect particle-level measurements to perturbative quantum-chromodynamic (QCD) calculations.   A model posits a mechanism for converting quarks and gluons (and possibly diquarks) into the observed hadrons, and is implemented through an algorithm and parametrized probability distributions.    Precision tests of the standard model at colliders rely on these hadronization models, which are embedded inside event generators.   It is of vital importance to test the validity of our hadronization models and to determine what range of parameters are consistent with data.   This is often a time-consuming task, because the models depend on many parameters, each of which must be propagated forward to make a comparison with data.

In this work, we consider the Lund string fragmentation model~\cite{Andersson:1983ia,Andersson:1997xwk}  as implemented in \textsc{Pythia} \cite{Bierlich:2022pfr}. 
Uncertainties on the parameters of this model often lead to significant uncertainties on theory predictions compared to data at colliders.
While the model can be more complex, we consider a simpler version and focus only on tuning kinematic parameters, which are assumed to be universal for all quark and diquark types. Flavor parameters are left unchanged from the current default values.

Hadron kinematics $(p_x,p_y,p_z,E)$ at each string break are governed by the sampling of two correlated distributions -- kinematics transverse to the string are sampled from a Gaussian
\begin{equation}\label{eq:pTGauss}
    \mathcal{P}(p_x, p_y; \sigma_{p_T}) = \frac{1}{2\pi \sigma_{p_T}} \exp \left( - \frac{p_x^2 + p_y^2}{2 \sigma_{p_T}^2} \right)\,,
\end{equation}
while kinematics longitudinal to the string motion are governed by the left-right Lund fragmentation function, defined up to a normalization constant as
\begin{equation}\label{eq:fragz}
    f(z) \propto \frac{(1-z)^a}{z} \exp \left( -\frac{b m_T^2}{z} \right)\,,
\end{equation}
where $\sigma_{p_T}, a,$ and $b$ are tuneable parameters fit to data, $z$ is the longitudinal momentum fraction defined as
\begin{equation}\label{eq:lightconez}
    z \equiv \frac{(E \pm p_z)_{\text{hadron}}}{(E \pm p_z)_{\text{string}}}\,,
\end{equation}
with $m_T^2 \equiv m_h^2 + p_T^2$.   The parameter $b$ sets the importance of different string areas, the parameter $a$ sets the fragmentation dependence as $z\to 1$, and $\sigma_{p_T}$ sets the range of $p_T$ values at each string break. Note that while the normalization factor cannot be calculated analytically, the maximum of $f(z)$ can be. For more details regarding the explicit implementation of the Lund model within \textsc{Pythia}, see ref.~\cite{Bierlich:2022pfr}.

\subsection{Fitting a hadronization model with RSA}\label{subsec:hadronization_fit_RSA}

To showcase the power of RSA, we focus on the two parameters which render rejection-based sampling necessary, $a$ and $b$, and keep $\sigma_{p_T}$ fixed to its default value, $0.335$. To fit the $a$ and $b$ parameters of the Lund fragmentation function using RSA requires a data-structure comprised of three components: (i) the fragmentation-chain-level rejection sampling data e.g.~the accepted and rejected longitudinal momentum fraction samples $\boldsymbol{z}$ as well as the transverse mass $m_T$ of the fragmenting hadron, see eq.~\ref{eq:fragz}; (ii) the desired measurable observables from simulated hadronization events produced by the base parameterization $\boldsymbol{y}_{\text{sim}}$ and (iii) values of the same observables measured experimentally $\boldsymbol{y}_{\text{exp}}$. 

Our simulation consists of the hadronization of a simple string system based on the probabilistic model described in Sec.~\ref{subsec:hadronization}.
An example of the training data, consisting of $N$ hadronization events, is therefore\footnote{While not denoted explicitly in eq.~\ref{eq:ARRG_training_data}, each array $\boldsymbol{z}_i$ is zero-padded to a fixed length.   In principle, the series of rejections can be stored with the random number seed used in the numerical algorithm.}
\begin{equation}
  \renewcommand\arraystretch{1.2}
  \boldsymbol{z} = \begin{pmatrix}
    \boldsymbol{z}_1 = \begin{pmatrix} \big\{ m_T^{h_1}, z^{h_1}_{\text{accept}}, z^{1, h_1}_{\text{reject}}, \cdots, 									  z^{n_{h_1}, h_1}_{\text{reject}}\big\} \\
    								   \big\{m_T^{h_2}, z^{h_2}_{\text{accept}}, z^{1, h_2}_{\text{reject}}, \cdots, z^{n_{h_2}, h_2}_{\text{reject}} \big\} \\	
    								   \big\{m_T^{h_3}, z^{h_3}_{\text{accept}}, z^{1, h_3}_{\text{reject}}, \cdots, z^{n_{h_3}, h_3}_{\text{reject}} \big\} \end{pmatrix}_1 \\ \\
    \boldsymbol{z}_2 = \begin{pmatrix} \big\{ m_T^{h_1}, z^{h_1}_{\text{accept}}, z^{1, h_1}_{\text{reject}}, \cdots, 									  z^{n_{h_1}, h_1}_{\text{reject}}\big\} \\
    								   \vdots \\
    								   \big\{m_T^{h_4}, z^{h_4}_{\text{accept}}, z^{1, h_4}_{\text{reject}}, \cdots, z^{n_{h_4}, h_4}_{\text{reject}} \big\} \end{pmatrix}_2 \\
    \vdots \\
    \boldsymbol{z}_N = \cdots
  \end{pmatrix},
  \label{eq:ARRG_training_data}
\end{equation}
where $n_{h_i}$ refers to the total number of rejections for the $i^\text{th}$ fragmentation/hadron. Note that a string system will fragment hadrons until its invariant mass reaches a predetermined threshold (usually set at $1$ GeV). When this threshold is reached, a different function is used to partition the remaining energy and momentum of the system between the final two hadrons such that energy and momentum is conserved and the left-right symmetry of the Lund model is preserved. This function, called \texttt{finalTwo}, can fail (roughly $10-15\%$ of the time for a $q\bar{q}$ string system with total energy of $50$ GeV).  Upon failure, the previous hadronization event is discarded and the hadronization routine is re-run on a reinitialized string. For this technical reason, the hadron multiplicity $N_h$ of an event is not necessarily equal to the total number of accepted $z$-values.
\begin{figure*}[t!]
  \centering
  \includegraphics[width=0.49\textwidth]{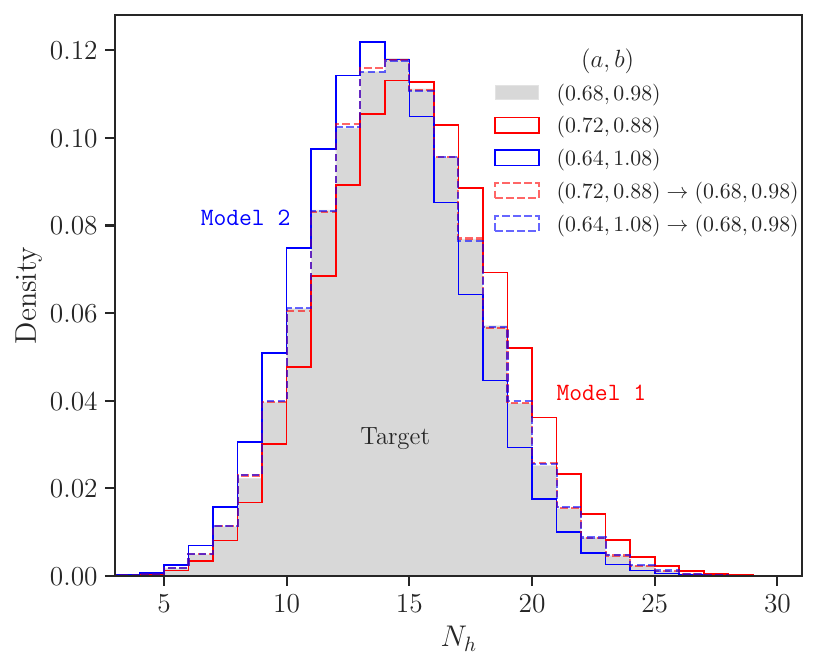}
  \includegraphics[width=0.469\textwidth]{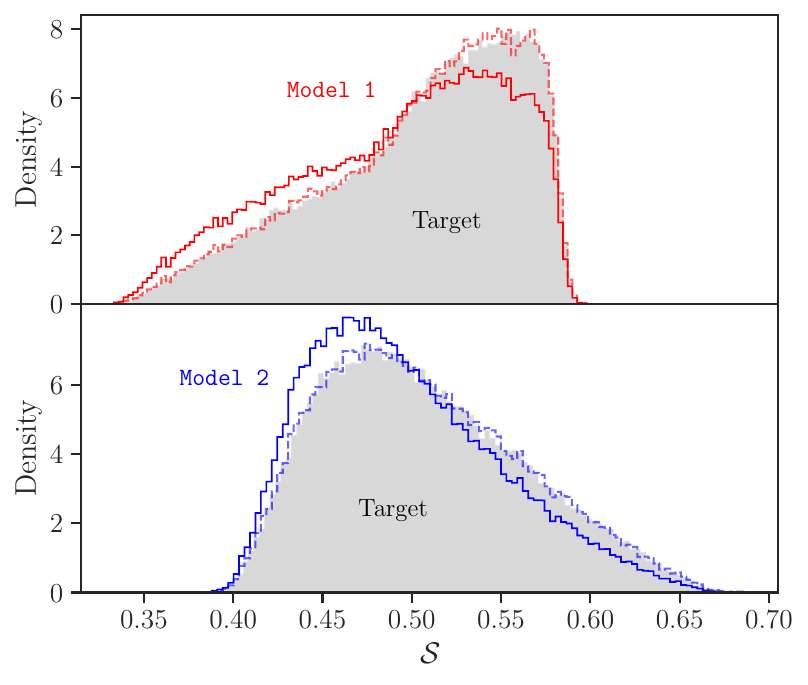}
  \caption{Observable distributions for hadron multiplicity $N_h$ (left) and classifier score $\mathcal{S}$ (right) at three distinct parameterizations of the Lund $a,b$ parameters. Dotted histograms denote the reweighed distributions.
  }
  \label{fig:observable_distributions}
\end{figure*}

Because RSA utilizes reweighting to explore the model parameter space, at least one set of events, generated using a base parameterization $\boldsymbol{\theta}_B \equiv \{a, b\}_B$, is required for tuning\footnote{When fitting to experimental data (or when a good guess for initial parameters is unknown) multiple base parameterizations would be desirable/required. For a detailed discussion see Sect.~\ref{sec:data}}. Given the unnormalized Lund fragmentation function, $f(z; \boldsymbol{\theta})$, the forward pass of training proceeds by first computing an event-weight array $\boldsymbol{w}$ for the proposed parameters, $\boldsymbol{\theta}_P \equiv \{a, b\}_P$, given by
\renewcommand\arraystretch{1.5}
\begin{equation}\label{eq:weight_array}
  \boldsymbol{w} = \begin{pmatrix}
    w_1 & w_2 & \cdots & w_N 
  \end{pmatrix}^T, \text{ where}
\end{equation}
\begin{align}\nonumber
  w_n =& \prod_{i = 1}^{\tilde{N}_{h,n}} \left(\frac{f(z^{h_i}_\text{accept}; \{a, b\}_P)}{f(z^{h_i}_\text{accept}; \{a, b\}_{B})}\right) \\\label{eq:frag_weights}
  &\hspace{8mm} \times \prod_{j = 1}^{n_{h_i}} \left(\frac{\hat{f} - f(z^{j, h_i}_\text{reject}; \{a, b\}_P)}{\hat{f} - f(z^{j, h_i}_\text{reject}; \{a, b\}_{B})}\right)\,,
\end{align}   
and $\hat{f}$ is the oversampling factor associated with the sampling of $z$ and $\tilde{N}_{h,n}$ denotes the the total number of distinct accept-reject samplings in the $n^\text{th}$ event (including those when \texttt{finalTwo} fails). Once the event-weight array has been computed, a `cost' or `loss' function must be defined and minimized to allow for the determination of the next set of parameters $\boldsymbol{\theta}_P$. The loss will parameterize the difference between $\boldsymbol{y}_{\text{exp}}$ and the weighted $\boldsymbol{y}_{\text{sim}}$ ensembles, see Sec.~\ref{sec:results}. Once the loss has been computed, PyTorch's automatic numerical differentiation engine \texttt{autograd} paired alongside a chosen optimization algorithm (\texttt{Adam}, \texttt{SGD}, $\ldots$ , etc.) can be used to choose the next model parameterization i.e. fragmentation function $f(z; \boldsymbol{\theta})$. 

\section{Data}\label{sec:data}

As a proof of principle for our method, we fit a simplified hadronizing system, namely a quark anti-quark pair hadronizing to pions ($q\bar{q} \rightarrow \pi$'s), using the Lund string model of hadronization. The events are generated from a `particle gun'\footnote{See, for example, \texttt{main234} within the \texttt{Pythia/examples} sub-directory of release $8.312$.} where the initial configuration is a $q\bar{q}$ system outgoing back-to-back oriented along the $z$-axis with each quark having equal energies $E$. We simulate $10^6$ of the specified hadronization events with $E = 50$ GeV using three parameterizations of the Lund fragmentation function: $\boldsymbol{\theta}_1 \equiv \textcolor{red}{\texttt{Model 1}} = \{a = 0.72, b = 0.88\}$, $\boldsymbol{\theta}_2 \equiv \textcolor{blue}{\texttt{Model 2}} = \{a = 0.64, b = 1.08\}$, and the \textsc{Pythia} default $\boldsymbol{\theta}_{\text{exp}} \equiv \text{Target} = \{a = 0.68, b = 0.98\}$. Both \textcolor{red}{\texttt{Model 1}} and \textcolor{blue}{\texttt{Model 2}} will be used as base parameterizations during the fits. The rejection sampling uses an over-sampling factor of $\hat{f} = 10$, each accept-reject array is zero-padded to a fixed length of 200\footnote{We do not allow for > 199 rejections. This is a convention chosen to reduce the memory footprint of the accept-reject datasets required for reweighting. We have checked empirically that this choice does not effect fit performance.}, and each event is zero-padded to a fixed length of 75. 

\begin{figure*}[th!]
  \centering
  \includegraphics[width=0.55\textwidth]{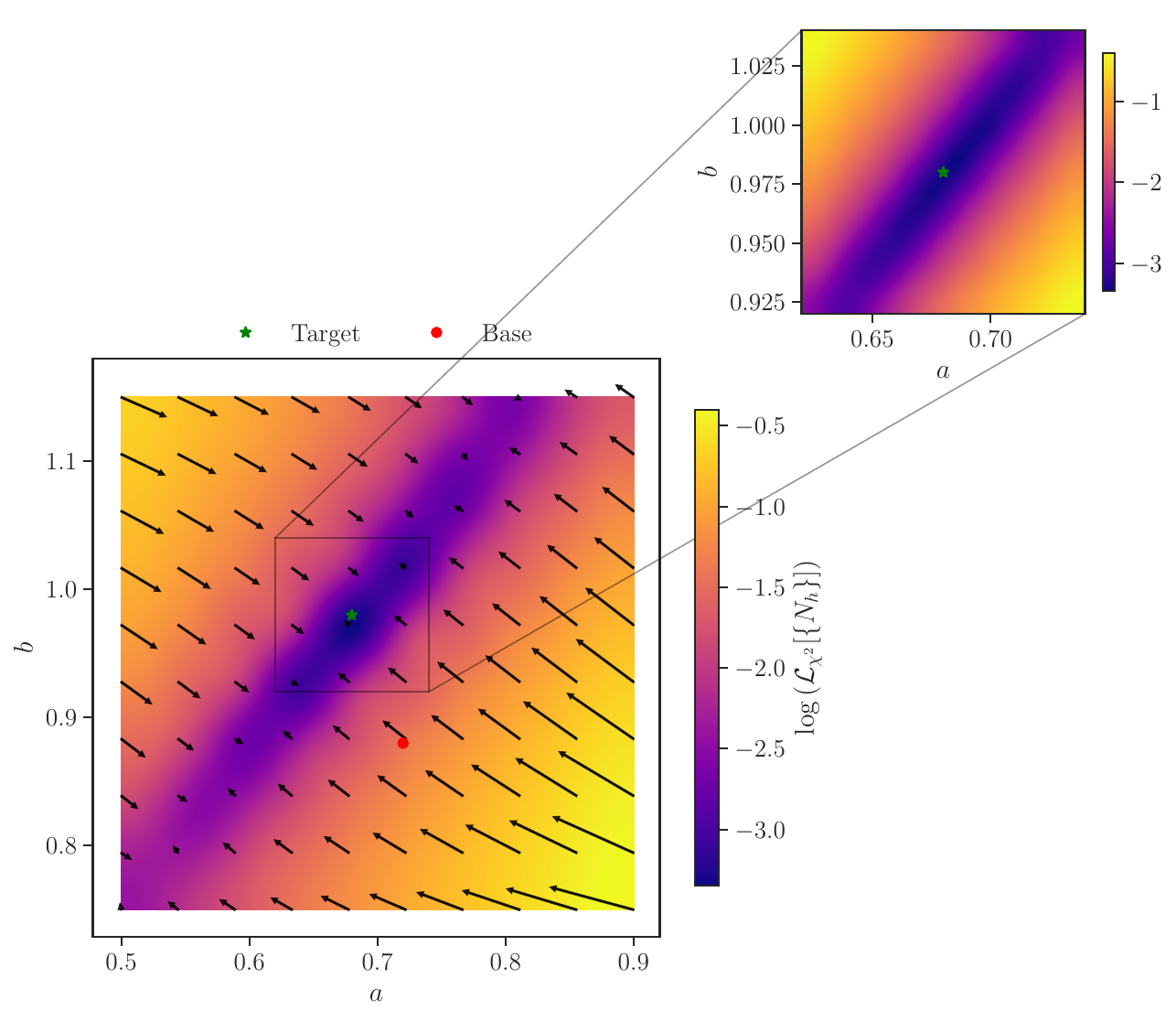}
  \caption{An example of the two-dimensional Lund $a$ and $b$ parameter plane loss landscape for the pseudo-$\chi^2$ loss defined in eq.~\ref{eq:chi2_loss}. The red dot denotes the initial base parameterization from which we reweigh from and the green cross denotes the target parameterization from which the observable distribution was generated from. Both subplots are generated by computing the loss and gradients over a grid of 100 $(a,b)$ pairs. For each parameterization the losses and gradients are averaged over four mini-batches of $5\times 10^{4}$ events for a total `batch size' of $2\times 10^5$ events.}
  \label{fig:loss_landscape_pseudo_chi2}
\end{figure*}

The efficacy/performance of a reweighting between two parameterizations can be summarized by two metrics
\begin{equation}\label{eq:metrics}
    \mu \equiv \sum_{i=1}^N \frac{w_i}{N}, \quad N_{\text{eff}} = \frac{\big(\sum_{i=1}^N w_i\big)^2}{\sum_{i=1}^N w_i^2}.
\end{equation}
The two metrics provide complementary reweighinting diagnostics -- the first gives a measure of possible lack of coverage or `unitarity violation' between the two parameterizations and the second reflects the possible loss in statistical power of the reweighed sample. For two parameterizations with sufficiently good coverage we expect $\mu \approx 1$. Likewise, $N_{\text{eff}}/N \approx 1$, with $N$ being the total number of samples used during reweighting, would signal very little statistical power lost to reweighting. In figure~\ref{fig:a_b_reweighting_metrics} we show both of these reweighting metrics across the $(a,b)$ parameter plane for both \textcolor{red}{\texttt{Model 1}} and \textcolor{blue}{\texttt{Model 2}}. We express $\mu$ (left column) as $\ln((1-\mu)/\sigma_{1-\mu})$, where $\sigma_{1-\mu}$ is the uncertainty on $1-\mu$, to better assess consistency with $\mu=1$. Written this way and for the sample sizes considered in this work, $\ln((1-\mu)/\sigma_{1-\mu})$ in the approximate range $(-1,1)$ are considered as reflecting adequate reweighting. In general, values closer to 0 imply better coverage, values $\gg 0$ are inconsistent with $0$ due to lack of coverage, and values $\ll 0$ indicate weights with large variance and signal poor reweighting performance. We see that the target parameterizations (green stars) sit in regions where the reweighting metrics indicate good coverage and effective statistics in both base models (red dots).
We also see that the full parameter range of interest contains large regions with poor coverage and effective statistics. When performing a black-box fit (where a good estimate of model parameters is unknown) over the full parameter range, one would need to utilize multiple base parameterizations that, when combined, provide adequate coverage over the full parameter volume.

\afterpage{
\begin{figure*}[p!]
  \centering
  \includegraphics[width=0.9\textwidth]{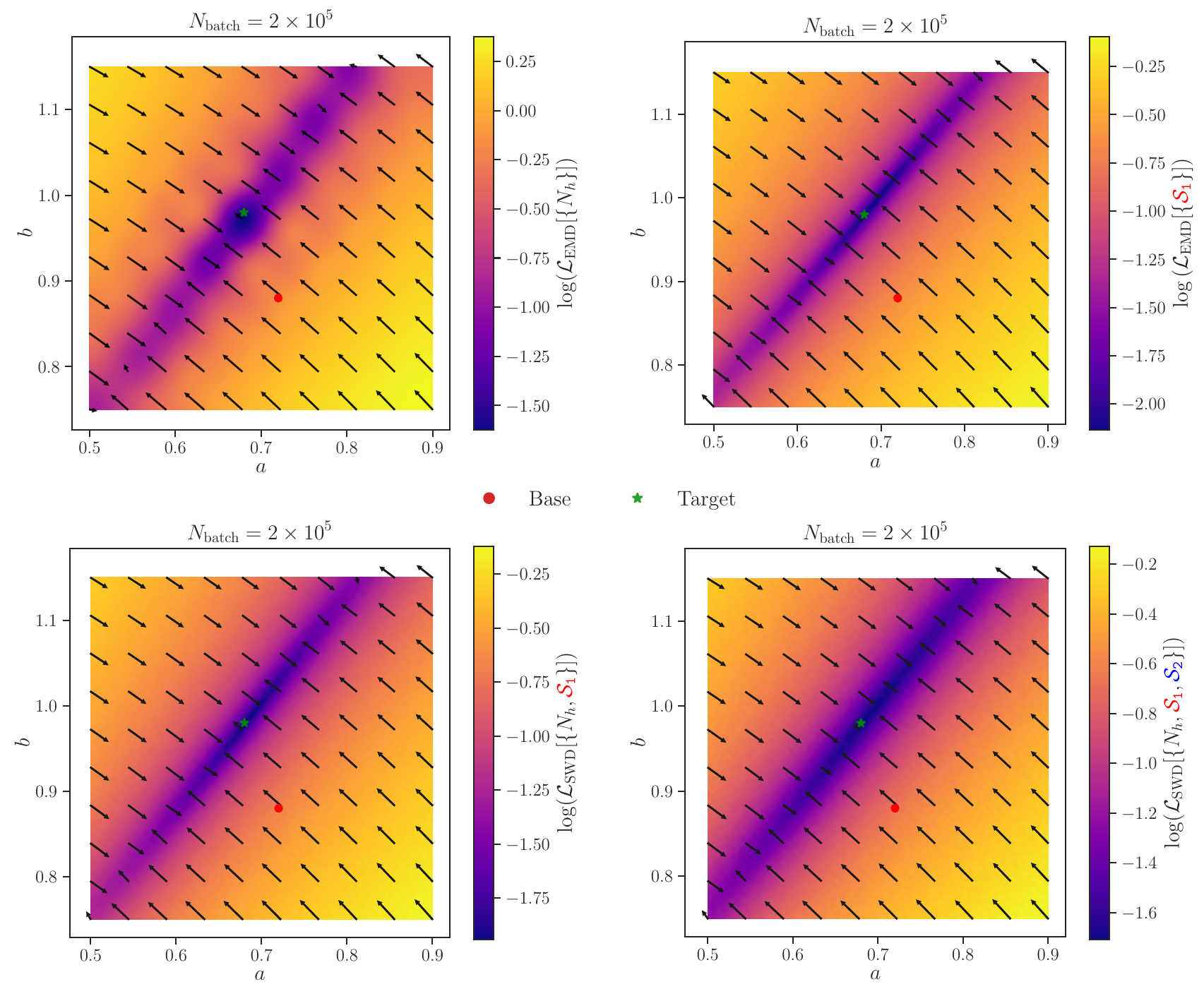}
  \caption{Two-dimensional Lund $a$ and $b$ parameter plane loss landscape for one-dimensional and sliced Wasserstein distances over various dimensionalities of low and high-level observables, with the same grid and mini-batching as in figure \ref{fig:loss_landscape_pseudo_chi2}.}
  \label{fig:loss_landscapes_OT}
\end{figure*}
\newpage
}

\subsection{Observables}\label{subsec:observables}
We categorize observables as either `high-level' or `low-level'.  High-level data utilizes currently available event-level/jet-level observables such as (charged) hadron multiplicity, event-shape variables like thrust, etc. 
Low-level data, on the other hand, assumes access to all hadron-level data that can in principle be extracted from experimental data, such as the energy and momenta of all hadrons within an event. 

For tuning, we consider two observables -- one high-level and one low-level.  The high-level observable is the hadron multiplicity $N_h$ that counts the total number of hadrons in the hadronization chain accepted by \texttt{finalTwo}. The multiplicity distributions are shown in the left panel of figure \ref{fig:observable_distributions} for the three model parameterizations as well as the reweighted distributions following eq.~\ref{eq:frag_weights}.
The second observable is the low-level \textit{truth score} $\mathcal{S}$. The truth score is the output of a trained DeepSets classifier~\cite{zaheer2018deepsets,Komiske:2018cqr} that distinguishes between full low-level input from simulated (base) and experimental (truth/target) data. 
The classifier consists of a per-particle multi-layer-perceptron (MLP) encoder $\phi$, an aggregation function chosen to be simple addition, and a MLP decoder $\rho$. Both the encoder and decoder networks consist of three layers and 64 hidden nodes. The input consists of low-level pion observables \eg~four-momenta that were preprocessed into transverse momentum, polar angle, and pseudorapidity $\mathbf{\Theta}^i_h \equiv (p_{T,i}, \phi_i, \eta_i)$ measured relative to the initial $z$-axis where
\begin{align}
    p_{T,i} &= \sqrt{p_{x,i}^2+p_{y,i}^2},\nonumber \\
    \phi_i &= \tan^{-1}(p_{y,i}/p_{x,i}), \nonumber \\
    \eta_i &= -\log\left[\tan^{-1}(p_z/2p_{T,i})\right].
\end{align}
The classifier was first trained with labeled data minimizing the binary-cross-entropy loss (BCE) between simulated and experimental data. After training, the score is given by
\begin{equation}
    \mathcal{S}(\mathbf{\Theta}^1_h, \cdots, \mathbf{\Theta}^{N_h}_h) = \rho\left(\sum_i^{N_h} \phi(\mathbf{\Theta}^i_h)\right).
\end{equation}
As an observable, the score $\mathcal{S}$ effectively compresses the full-phase space kinematics $\left\{\mathbf{\Theta}^1_h,\cdots,\mathbf{\Theta}^{N_h}_h\right\}$ of all $N_h$ hadrons in an event into a single scalar representation with hadron permutation-invariance that can be applied to any dataset irrespective of the actual parameter values considered for generation. The right panel of figure \ref{fig:observable_distributions} shows the score distributions as well as the reweighted distributions for the two cases in which the classifier is initially trained to distinguish between \textcolor{red}{\texttt{Model 1}} (right top) or \textcolor{blue}{\texttt{Model 2}} (right bottom) from Target which we will refer to as scores \textcolor{red}{$\mathcal{S}_{\texttt{1}}$} and \textcolor{blue}{$\mathcal{S}_{\texttt{2}}$}, respectively.
Finally, when fitting, we consider multiplicity and scores individually as one-dimensional distributions as well as together to form two- and three-dimensional distributions, e.g.
\begin{equation}
\boldsymbol{y}^{d}_\text{sim} = \begin{pmatrix}
    \boldsymbol{y}_1 = 
     C(\{N_{h,1},\textcolor{red}{\mathcal{S}_{\texttt{1},1}},\textcolor{blue}{\mathcal{S}_{\texttt{2},1}}\}, d ) \\
    \vdots \\
    \boldsymbol{y}_N = C\left(\{N_{h,N},\textcolor{red}{\mathcal{S}_{\texttt{1},N}},\textcolor{blue}{\mathcal{S}_{\texttt{2},N}}\}, d \right)
  \end{pmatrix}
\end{equation}
where $C(\boldsymbol{O},d)$ implies $\boldsymbol{O}$ choose $d$, with $d=1,2$ or 3 the number of observables considered, and keeping the same choice for all $\{\boldsymbol{y}^d_{\text{sim}}\}_{n=1}^{N}$.

\section{Results}\label{sec:results}
\begin{figure*}[t!]
    \centering
    \includegraphics[width=0.49\textwidth]{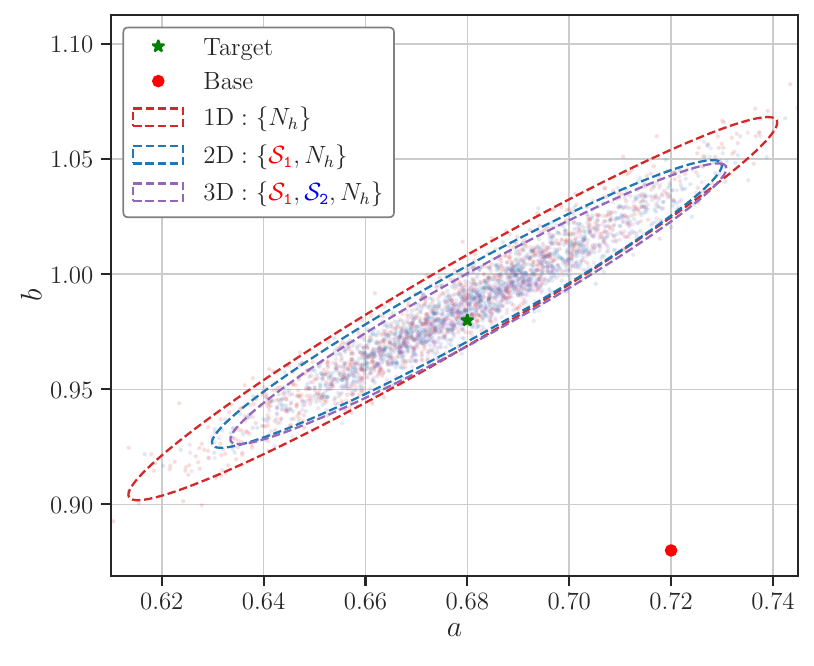}
    \caption{The 95\% confidence ellipses on $a$, $b$ parameters estimated from 1,000 bootstrapped training runs over 300 epochs and identical tuning hyperparameters. Demonstrates the increasing accuracy of slicing over multiple low and high-level observables, from only $N_h$ (red), $\{\textcolor{red}{\mathcal{S}_{\texttt{1}}}, N_h\}$ (blue), and $\{\textcolor{red}{\mathcal{S}_{\texttt{1}}}, \textcolor{blue}{\mathcal{S}_{\texttt{2}}}, N_h\}$ (purple).}
    \label{fig:bootstrap}
\end{figure*}

The algorithmic prescription for parameter estimation using RSA described in Sec.~\ref{sec:tuning_w_RSA} offers an efficient method for analyzing and exploring model parameter space. 
In addition to the standard binned and unbinned observables used in modern tuning exercises, RSA facilitates the exploration of parameter space using \textit{machine-learning-based observables} (such as the score $\mathcal{S}$). This facilitates a systematic study and comparison of metric spaces utilizing high- versus low-level observables. Here we first validate the algorithm through the construction of loss landscapes for a variety of metrics and then turn to the investigation of fit performance using high- versus low-level observables.
\subsection{Loss landscapes}\label{subsec:loss_landscapes}
Given an accept-reject data-sample for an arbitrary parameterization $\boldsymbol{\theta}$, as described in Sec.~\ref{subsec:hadronization_fit_RSA}, and a set of observables $\boldsymbol{y}_{\text{exp}}, \boldsymbol{y}_{\text{sim}}$, RSA with reweighting can be used as a powerful exploratory device over the parameter volume. Once a metric comparing $\boldsymbol{y}_{\text{exp}}$ and $\boldsymbol{y}_{\text{sim}}$ is defined, a `loss landscape' that illustrates the magnitude and gradients of the metric over $\boldsymbol{\theta}$ can be created quickly using reweighting. Below we construct loss landscapes for various losses and observables.
\paragraph{Pseudo-$\chi^2$}
For binned observables, a commonly used metric to determine fit efficacy is the pseudo-$\chi^2$ \cite{Skands:2014pea,Ilten:2016csi}. Given a differentiable binning of $\boldsymbol{y}_{\text{sim}}, \boldsymbol{y}_{\text{exp}}$ and weights $\boldsymbol{w}$ the pseudo-$\chi^2$ loss can be expressed as:
\begin{equation}\label{eq:chi2_loss}
	\mathcal{L}_{\chi^2}(\boldsymbol{y}_{\text{sim}}, \boldsymbol{y}_{\text{exp}}; \boldsymbol{w}) = \sum_{i = 1}^{n_{\text{bins}}} \frac{(y^{(i)}_\text{sim} - y_\text{exp}^{(i)})^2}{\sigma^2_{\text{sim},i} + \sigma^2_{\text{exp},i}}
\end{equation}
where $y^{(i)}$ represents the normalized count density in the $i^\text{th}$ bin and $\sigma$ denotes the statistical uncertainty of the $i^\text{th}$ simulated and experimental bin. Note that this loss is well-defined for arbitrary multi-dimensional observable distributions. The landscape of the  pseudo-$\chi^2$ for multiplicity $N_h$ distributions over the parameter plane can be seen in figure \ref{fig:loss_landscape_pseudo_chi2}.   

\paragraph{1D Wasserstein distance}
For unbinned observables, we utilize the optimal-transport-based Wasserstein, or Earth mover's distance (EMD). Unlike $f$-divergences (like the $\chi^2$), the EMD does not require overlapping support.  Tuning with the EMD is similar to a WGAN~\cite{arjovsky2017wassersteingan}, only here, the EMD is computed explicitly and thus there is no trainable `critic' (making the optimization more stable)\footnote{A GAN-based reweighting and fitting has been explored in ref.~\cite{Desai:2024kpd}.}.
A similar use of the EMD is found in ref.~\cite{Pan:2024rfh}, which used this setup to learn surrogate models for likelihood ratios.
The EMD loss is defined as
\begin{equation}\label{eq:EMD_loss}
  \llh[\emd](\bysim, \byexp; \boldsymbol{w}) =
  \sum_{n = 1}^N\sum_{m = 1}^M f^*_{n,m} d_{n,m}\mmp{,}
\end{equation}
where the elements of the flow matrix $f^*_{n,m}$ encode the fractional amount of weight to be transferred between event $\bysim[,n]$ and $\byexp[,m]$ and $d_{n,m}=||\bysim[,n]-\byexp[,m]||_{2}$ is the distance between these two events. Here, $M$ is the number of events observed in the experimental dataset. The top panel of figure \ref{fig:loss_landscapes_OT} shows the EMD loss landscapes for both the multiplicity $N_h$ (top left) and the score $\textcolor{red}{\mathcal{S}_{\texttt{1}}}$ (top right). 

\paragraph{Multi-dimensional sliced Wasserstein distance}
To fit across multiple observables, such as multiplicity and classifier scores, we employ the sliced Wasserstein distance \cite{rabin2012wasserstein,kolouri2019generalized}. 
The observable vector is randomly projected onto a large number $N_P$ of random vectors $\mathbf{\theta}_i$ over the full observable space:
\begin{equation}\label{eq:sliced_EMD_loss}
  \llh[\text{SWD}](\bysim, \byexp; \boldsymbol{w}) =
  \frac{1}{N_P} \sum_{i = 1}^{N_P} \llh[\emd]({\by_{\text{sim},\theta_i}\xspace}, {\by_{\text{exp},\theta_i}\xspace}; \boldsymbol{w}).
\end{equation}
We apply slicing for two and three dimensional combinations of both low- and and high-level observables, over one classifier score and multiplicity as well as over two classifier scores based of two $(a,b)$ parameterizations and multiplicity. We chose $N_P = 128$ (with little sensitivity to small changes) and prescaled each observable to have zero mean and unit variance. The bottom panel of figure \ref{fig:loss_landscapes_OT} shows the sliced Wasserstein loss landscapes for both the 2D observable combining multiplicity with the \textcolor{red}{\texttt{Model 1}} score $\{N_h, \textcolor{red}{\mathcal{S}_{\texttt{1}}}\}$ (bottom left) and the 3D observable combining multiplicity with the \textcolor{red}{\texttt{Model 1}} and \textcolor{blue}{\texttt{Model 2}} scores $\{N_h,\textcolor{red}{\mathcal{S}_{\texttt{1}}}, \textcolor{blue}{\mathcal{S}_{\texttt{2}}}\}$ (bottom right). 

Note that in all cases (for all losses and observables), the loss landscapes in figures \ref{fig:loss_landscape_pseudo_chi2} and \ref{fig:loss_landscapes_OT} reveal a valley in the ($a,b$) plane illuminating an unavoidable degeneracy in model parameter space and correlation between the $a$ and $b$ parameters. For more details see the discussions in Refs. \cite{Ilten:2016csi,Skands:2014pea}.

\subsection{Confidence Intervals}

To demonstrate that the classifier score observables $\textcolor{red}{\mathcal{S}_{\texttt{1}}}$ and $\textcolor{blue}{\mathcal{S}_{\texttt{2}}}$ have learned to distinguish data based off of more than the high-level event multiplicity alone, we apply our Wasserstein tuning on the 1D, 2D, and 3D observables mentioned in Sec.~\ref{subsec:observables} and compare their performance. For each observable we perform 1,000 independent tunes, each using a single bootstrapped sample~\cite{10.1214/aos/1176344552} of 25,000 events (both from the target and \textcolor{red}{\texttt{Model 1}} parameterizations) randomly sampled from the full datasets. Each run is initialized at $\boldsymbol{\theta}_\text{init} = \textcolor{red}{\texttt{Model 1}}$ with the same training hyperparameters and trained over 300 epochs with the same bootstrapped dataset after each parameter update using the \texttt{AdamW} optimizer. In figure \ref{fig:bootstrap} we show the 95\% confidence ellipses over all bootstrapped tunes in the ($a,b$) plane. The one-dimensional observable tunes utilize the 1D Wasserstein loss while the higher-dimensional observables utilize the sliced Wasserstein loss as described in Sec.~\ref{subsec:loss_landscapes}.
We observe that the higher-dimensional observables (1D versus 2D versus 3D) produce a steady narrowing of confidence ellipses, indicating better fit performance.

\section{Conclusions and Outlook}\label{sec:conclusion}

In this paper we investigated the implications and applications of pairing rejection sampling with modern autodifferentiation engines. In particular, we focused in the context of model parameter estimation via reweighting and validated the methodology using a hadronization model. We developed an algorithmic prescription for RSA-based parameter estimation and showed that the methodology provides an efficient tool for model parameter space exploration. An important implication of embedding rejection sampling into the computational framework of differential libraries is the possibility of utilizing machine-learning-based observables during tuning. By comparing observables utilizing high-level event information, such as multiplicity, with those utilizing all available event-level information, like the classifier score, we showed that ML-based observables can improve fit performance and convergence.

There are two caveats to RSA-based parameter fits. Firstly, as discussed in Sec.~\ref{sec:data}, due to the degraded performance of weighted distributions far from the base parameterization, full black-box tuning exercises will likely require multiple base parameterizations to ensure proper coverage over the full parameter space. In practice, however, the total number of base parameterization can be minimized through successive coarse graining of the loss landscapes. Secondly, while not a fundamental issue, the memory footprint of the accept-reject datasets, due to the necessity of zero-padding, can become computationally prohibitive for large oversampling factors or high multiplicity final states (large string energies). This can be mitigated through oversampling factor tuning, seed storage, or the direct computation and in-place storage of gradients during event generation which would completely eliminate the need for external datasets all-together.  

While we performed a two-parameter fit for simplicity, modern differential programming libraries and auto-differentiation engines are commonly used to store and compute numerical gradient information on hundreds to millions of parameters. Because of this, we do not expect that the addition of more tuning parameters will admit any fundamental issues to the numerical algorithm itself (although with more parameters the choice of non-degenerate observables and high-fidelity loss functions becomes extremely non-trivial). In the context of hadronization models, applying the methods developed here in the flavor sector, where there are $\mathcal{O}(10^2)$ of tuneable parameters, may prove to be a good testing ground for the capacity of these methods. 

Tangentially, RSA methods may also have applications in numerical uncertainty quantification. As discussed in ref.~\cite{Bierlich:2023fmh}, reweighting can be useful for estimating modeling uncertainties. With RSA, direct access to numerical gradient information can provide a more rigorous characterization of model uncertainties by facilitating the computation of covariance matrices on fit parameters or the construction of maximum likelihood test statistics. We leave these considerations for future work.

RSA presents a versatile and flexible approach to parameter estimation. The methods developed here serve as a promising foundation for the next generation of simulation-based parameter estimation, offering a scalable approach to accurate, data-driven model tuning.

\section{Acknowledgments}
NH and BN are supported by the U.S. Department of Energy (DOE), Office of Science under contract DE-AC02-05CH11231.
PI is also supported by NSF grant NSF-PHY-2209769.
PI, TM, SM, MS, and AY acknowledge support in part by NSF grants OAC-2103889, OAC-2411215 and OAC-2417682. 
SM is supported by Fermi Research Alliance, LLC under Contract No. DE-AC02-07CH11359 with the U.S. Department of Energy, Office of Science, Office of High Energy Physics.
TM, MS, and AY are also in part funded by DOE grant DE-SC101977.
AS is supported by grant 2019/34/E/ST2/00457 of the National Science Centre, Poland and by the Priority Research Area Digiworld under the program `Excellence Initiative -- Research University' at the Jagiellonian University in Krakow. 
AY acknowledges support in part by The University of Cincinnati URC Graduate Support Program.  

\vspace{0.3in}
\noindent \textbf{Public code}: The open-source software developed for this project is publicly available at \url{https://github.com/tonymenzo/RSA}. Corresponding datasets are accessible at \url{https://zenodo.org/records/14289503}. All data was generated using \textsc{Pythia} (\texttt{8.312}), and the codebase was implemented in \texttt{Python} (\texttt{v3.12}) with \texttt{PyTorch} (\texttt{v2.5}).

\newpage
\bibliography{RSA}

\end{document}